


%
\expandafter\ifx\csname phyzzx\endcsname\relax\else
 \errhelp{Hit <CR> and go ahead.}
 \errmessage{PHYZZX macros are already loaded or input. }
 \endinput \fi
\catcode`\@=11 
%
%
%
\font\seventeenrm=cmr17
\font\fourteenrm=cmr12 scaled\magstep1
\font\twelverm=cmr12
\font\ninerm=cmr9            \font\sixrm=cmr6
%
\font\fourteenbf=cmbx10 scaled\magstep2
\font\twelvebf=cmbx12
\font\ninebf=cmbx9            \font\sixbf=cmbx6
%
\font\fourteeni=cmmi10 scaled\magstep2      \skewchar\fourteeni='177
\font\twelvei=cmmi12			        \skewchar\twelvei='177
\font\ninei=cmmi9                           \skewchar\ninei='177
\font\sixi=cmmi6                            \skewchar\sixi='177
%
\font\fourteensy=cmsy10 scaled\magstep2     \skewchar\fourteensy='60
\font\twelvesy=cmsy10 scaled\magstep1	    \skewchar\twelvesy='60
\font\ninesy=cmsy9                          \skewchar\ninesy='60
\font\sixsy=cmsy6                           \skewchar\sixsy='60
%
\font\fourteenex=cmex10 scaled\magstep2
\font\twelveex=cmex10 scaled\magstep1
%
\font\fourteensl=cmsl12 scaled\magstep1
\font\twelvesl=cmsl12
\font\ninesl=cmsl9
%
\font\fourteenit=cmti12 scaled\magstep1
\font\twelveit=cmti12
\font\nineit=cmti9
\font\fourteentt=cmtt10 scaled\magstep2
\font\twelvett=cmtt12
\font\fourteencp=cmcsc10 scaled\magstep2
\font\twelvecp=cmcsc10 scaled\magstep1
\font\tencp=cmcsc10
\newfam\cpfam
\newdimen\b@gheight		\b@gheight=12pt
\newcount\f@ntkey		\f@ntkey=0
\def\f@m{\afterassignment\samef@nt\f@ntkey=}
\def\samef@nt{\fam=\f@ntkey \the\textfont\f@ntkey\relax}
\def\rm{\f@m0 }
\def\mit{\f@m1 }         
\def\cal{\f@m2 }
\def\it{\f@m\itfam}
\def\sl{\f@m\slfam}
\def\bf{\f@m\bffam}
\def\tt{\f@m\ttfam}
\def\caps{\f@m\cpfam}
\def\fourteenpoint{\relax
    \textfont0=\fourteenrm          \scriptfont0=\tenrm
      \scriptscriptfont0=\sevenrm
    \textfont1=\fourteeni           \scriptfont1=\teni
      \scriptscriptfont1=\seveni
    \textfont2=\fourteensy          \scriptfont2=\tensy
      \scriptscriptfont2=\sevensy
    \textfont3=\fourteenex          \scriptfont3=\twelveex
      \scriptscriptfont3=\tenex
    \textfont\itfam=\fourteenit     \scriptfont\itfam=\tenit
    \textfont\slfam=\fourteensl     \scriptfont\slfam=\tensl
    \textfont\bffam=\fourteenbf     \scriptfont\bffam=\tenbf
      \scriptscriptfont\bffam=\sevenbf
    \textfont\ttfam=\fourteentt
    \textfont\cpfam=\fourteencp
    \samef@nt
    \b@gheight=14pt
    \setbox\strutbox=\hbox{\vrule height 0.85\b@gheight
				depth 0.35\b@gheight width\z@ }}
\def\twelvepoint{\relax
    \textfont0=\twelverm          \scriptfont0=\ninerm
      \scriptscriptfont0=\sixrm
    \textfont1=\twelvei           \scriptfont1=\ninei
      \scriptscriptfont1=\sixi
    \textfont2=\twelvesy           \scriptfont2=\ninesy
      \scriptscriptfont2=\sixsy
    \textfont3=\twelveex          \scriptfont3=\tenex
      \scriptscriptfont3=\tenex
    \textfont\itfam=\twelveit     \scriptfont\itfam=\nineit
    \textfont\slfam=\twelvesl     \scriptfont\slfam=\ninesl
    \textfont\bffam=\twelvebf     \scriptfont\bffam=\ninebf
      \scriptscriptfont\bffam=\sixbf
    \textfont\ttfam=\twelvett
    \textfont\cpfam=\twelvecp
    \samef@nt
    \b@gheight=12pt
    \setbox\strutbox=\hbox{\vrule height 0.85\b@gheight
				depth 0.35\b@gheight width\z@ }}
\def\tenpoint{\relax
    \textfont0=\tenrm          \scriptfont0=\sevenrm
      \scriptscriptfont0=\fiverm
    \textfont1=\teni           \scriptfont1=\seveni
      \scriptscriptfont1=\fivei
    \textfont2=\tensy          \scriptfont2=\sevensy
      \scriptscriptfont2=\fivesy
    \textfont3=\tenex          \scriptfont3=\tenex
      \scriptscriptfont3=\tenex
    \textfont\itfam=\tenit     \scriptfont\itfam=\seveni
    \textfont\slfam=\tensl     \scriptfont\slfam=\sevenrm
    \textfont\bffam=\tenbf     \scriptfont\bffam=\sevenbf
      \scriptscriptfont\bffam=\fivebf
    \textfont\ttfam=\tentt
    \textfont\cpfam=\tencp
    \samef@nt
    \b@gheight=10pt
    \setbox\strutbox=\hbox{\vrule height 0.85\b@gheight
				depth 0.35\b@gheight width\z@ }}
%
%
%
\normalbaselineskip = 20pt plus 0.2pt minus 0.1pt
\normallineskip = 1.5pt plus 0.1pt minus 0.1pt
\normallineskiplimit = 1.5pt
\newskip\normaldisplayskip
\normaldisplayskip = 20pt plus 5pt minus 10pt
\newskip\normaldispshortskip
\normaldispshortskip = 6pt plus 5pt
\newskip\normalparskip
\normalparskip = 6pt plus 2pt minus 1pt
\newskip\skipregister
\skipregister = 5pt plus 2pt minus 1.5pt
\newif\ifsingl@    \newif\ifdoubl@
\newif\iftwelv@    \twelv@true
\def\singlespace{\singl@true\doubl@false\spaces@t}
\def\doublespace{\singl@false\doubl@true\spaces@t}
\def\normalspace{\singl@false\doubl@false\spaces@t}
\def\Tenpoint{\tenpoint\twelv@false\spaces@t}
\def\Twelvepoint{\twelvepoint\twelv@true\spaces@t}
\def\spaces@t{\relax
      \iftwelv@ \ifsingl@\subspaces@t3:4;\else\subspaces@t1:1;\fi
       \else \ifsingl@\subspaces@t3:5;\else\subspaces@t4:5;\fi \fi
      \ifdoubl@ \multiply\baselineskip by 5
         \divide\baselineskip by 4 \fi }
\def\subspaces@t#1:#2;{
      \baselineskip = \normalbaselineskip
      \multiply\baselineskip by #1 \divide\baselineskip by #2
      \lineskip = \normallineskip
      \multiply\lineskip by #1 \divide\lineskip by #2
      \lineskiplimit = \normallineskiplimit
      \multiply\lineskiplimit by #1 \divide\lineskiplimit by #2
      \parskip = \normalparskip
      \multiply\parskip by #1 \divide\parskip by #2
      \abovedisplayskip = \normaldisplayskip
      \multiply\abovedisplayskip by #1 \divide\abovedisplayskip by #2
      \belowdisplayskip = \abovedisplayskip
      \abovedisplayshortskip = \normaldispshortskip
      \multiply\abovedisplayshortskip by #1
        \divide\abovedisplayshortskip by #2
      \belowdisplayshortskip = \abovedisplayshortskip
      \advance\belowdisplayshortskip by \belowdisplayskip
      \divide\belowdisplayshortskip by 2
      \smallskipamount = \skipregister
      \multiply\smallskipamount by #1 \divide\smallskipamount by #2
      \medskipamount = \smallskipamount \multiply\medskipamount by 2
      \bigskipamount = \smallskipamount \multiply\bigskipamount by 4 }
\def\normalbaselines{ \baselineskip=\normalbaselineskip
   \lineskip=\normallineskip \lineskiplimit=\normallineskip
   \iftwelv@\else \multiply\baselineskip by 4 \divide\baselineskip by 5
     \multiply\lineskiplimit by 4 \divide\lineskiplimit by 5
     \multiply\lineskip by 4 \divide\lineskip by 5 \fi }
\Twelvepoint  
\interlinepenalty=50
\interfootnotelinepenalty=5000
\predisplaypenalty=9000
\postdisplaypenalty=500
\hfuzz=1pt
\vfuzz=0.2pt
\voffset=0pt
\dimen\footins=8 truein
%
%
%
\def\pagecontents{
   \ifvoid\topins\else\unvbox\topins\vskip\skip\topins\fi
   \dimen@ = \dp255 \unvbox255
   \ifvoid\footins\else\vskip\skip\footins\footrule\unvbox\footins\fi
   \ifr@ggedbottom \kern-\dimen@ \vfil \fi }
\def\makeheadline{\vbox to 0pt{ \skip@=\topskip
      \advance\skip@ by -12pt \advance\skip@ by -2\normalbaselineskip
      \vskip\skip@ \line{\vbox to 12pt{}\the\headline} \vss
      }\nointerlineskip}
\def\makefootline{\baselineskip = 1.5\normalbaselineskip
                 \line{\the\footline}}
\newif\iffrontpage
\newif\ifletterstyle
\newif\ifp@genum
\def\nopagenumbers{\p@genumfalse}
\def\pagenumbers{\p@genumtrue}
\pagenumbers
\newtoks\paperheadline
\newtoks\letterheadline
\newtoks\paperfootline
\newtoks\letterfootline
\newtoks\letterinfo
\newtoks\Letterinfo
\newtoks\date
\footline={\ifletterstyle\the\letterfootline\else\the\paperfootline\fi}
\paperfootline={\hss\iffrontpage\else\ifp@genum\tenrm\folio\hss\fi\fi}
\letterfootline={\iffrontpage\LETTERFOOT\else\hfil\fi}
\Letterinfo={\hfil}
\letterinfo={\hfil}
\def\LETTERFOOT{\hfil} 
%
\def\LETTERHEAD{\vtop{\baselineskip=20pt\hbox to
\hsize{\hfil\seventeenrm\strut
CALIFORNIA INSTITUTE OF TECHNOLOGY \hfil}
\hbox to \hsize{\hfil\ninerm\strut
CHARLES C. LAURITSEN LABORATORY OF HIGH ENERGY PHYSICS \hfil}
\hbox to \hsize{\hfil\ninerm\strut
PASADENA, CALIFORNIA 91125 \hfil}}}
\headline={\ifletterstyle\the\letterheadline\else\the\paperheadline\fi}
\paperheadline={\hfil}
\letterheadline{\iffrontpage \LETTERHEAD\else
    \rm \ifp@genum \hfil \folio\hfil\fi\fi}
\def\monthname{\relax\ifcase\month 0/\or January\or February\or
   March\or April\or May\or June\or July\or August\or September\or
   October\or November\or December\else\number\month/\fi}
\def\today{\monthname\ \number\day, \number\year}
\date={\today}
\countdef\pageno=1      \countdef\pagen@=0
\countdef\pagenumber=1  \pagenumber=1
\def\advancepageno{\global\advance\pagen@ by 1
   \ifnum\pagenumber<0 \global\advance\pagenumber by -1
    \else\global\advance\pagenumber by 1 \fi \global\frontpagefalse }
\def\folio{\ifnum\pagenumber<0 \romannumeral-\pagenumber
           \else \number\pagenumber \fi }
\def\footrule{\dimen@=\prevdepth\nointerlineskip
   \vbox to 0pt{\vskip -0.25\baselineskip \hrule width 0.35\hsize \vss}
   \prevdepth=\dimen@ }
\newtoks\foottokens
\foottokens={}
\newdimen\footindent
\footindent=24pt
\def\vfootnote#1{\insert\footins\bgroup
   \interlinepenalty=\interfootnotelinepenalty \floatingpenalty=20000
   \singl@true\doubl@false\Tenpoint
   \splittopskip=\ht\strutbox \boxmaxdepth=\dp\strutbox
   \leftskip=\footindent \rightskip=\z@skip
   \parindent=0.5\footindent \parfillskip=0pt plus 1fil
   \spaceskip=\z@skip \xspaceskip=\z@skip
   \the\foottokens
   \Textindent{$ #1 $}\footstrut\futurelet\next\fo@t}
\def\Textindent#1{\noindent\llap{#1\enspace}\ignorespaces}
\def\footnote#1{\attach{#1}\vfootnote{#1}}

\def\foot{\attach\footsymbolgen\vfootnote{\footsymbol}}
\let\footsymbol=\star
\newcount\lastf@@t           \lastf@@t=-1
\newcount\footsymbolcount    \footsymbolcount=0
\newif\ifPhysRev
\def\footsymbolgen{\bumpfootsymbolcount \generatefootsymbol \footsymbol }
\def\bumpfootsymbolcount{\relax
   \iffrontpage \bumpfootsymbolNP \else \advance\lastf@@t by 1
     \ifPhysRev \bumpfootsymbolPR \else \bumpfootsymbolNP \fi \fi
   \global\lastf@@t=\pagen@ }
\def\bumpfootsymbolNP{\ifnum\footsymbolcount <0 \global\footsymbolcount =0 \fi
    \ifnum\lastf@@t<\pagen@ \global\footsymbolcount=0
     \else \global\advance\footsymbolcount by 1 \fi }
\def\bumpfootsymbolPR{\ifnum\footsymbolcount >0 \global\footsymbolcount =0 \fi
      \global\advance\footsymbolcount by -1 }
\def\fd@f#1 {\xdef\footsymbol{\mathchar"#1 }}
\def\generatefootsymbol{\ifcase\footsymbolcount \fd@f 13F \or \fd@f 279
	\or \fd@f 27A \or \fd@f 278 \or \fd@f 27B \else
	\ifnum\footsymbolcount <0 \fd@f{023 \number-\footsymbolcount }
	 \else \fd@f 203 {\loop \ifnum\footsymbolcount >5
		\fd@f{203 \footsymbol } \advance\footsymbolcount by -1
		\repeat }\fi \fi }

\def\nonfrenchspacing{\sfcode`\.=3001 \sfcode`\!=3000 \sfcode`\?=3000
	\sfcode`\:=2000 \sfcode`\;=1500 \sfcode`\,=1251 }
\nonfrenchspacing
\newdimen\d@twidth
{\setbox0=\hbox{s.} \global\d@twidth=\wd0 \setbox0=\hbox{s}
	\global\advance\d@twidth by -\wd0 }
\def\removehglue{\loop \unskip \ifdim\lastskip >\z@ \repeat }
\def\roll@ver#1{\removehglue \nobreak \count255 =\spacefactor \dimen@=\z@
	\ifnum\count255 =3001 \dimen@=\d@twidth \fi
	\ifnum\count255 =1251 \dimen@=\d@twidth \fi
    \iftwelv@ \kern-\dimen@ \else \kern-0.83\dimen@ \fi
   #1\spacefactor=\count255 }
\def\step@ver#1{\relax \ifmmode #1\else \ifhmode
	\roll@ver{${}#1$}\else {\setbox0=\hbox{${}#1$}}\fi\fi }
\def\attach#1{\step@ver{\strut^{\mkern 2mu #1} }}
%
%
%
\newcount\chapternumber      \chapternumber=0
\newcount\sectionnumber      \sectionnumber=0
\newcount\equanumber         \equanumber=0
\let\chapterlabel=\relax
\let\sectionlabel=\relax
\newtoks\chapterstyle        \chapterstyle={\Number}
\newtoks\sectionstyle        \sectionstyle={\chapterlabel\Number}
\newskip\chapterskip         \chapterskip=\bigskipamount
\newskip\sectionskip         \sectionskip=\medskipamount
\newskip\headskip            \headskip=8pt plus 3pt minus 3pt
\newdimen\chapterminspace    \chapterminspace=15pc
\newdimen\sectionminspace    \sectionminspace=10pc
\newdimen\referenceminspace  \referenceminspace=25pc
\def\chapterreset{\global\advance\chapternumber by 1
   \ifnum\equanumber<0 \else\global\equanumber=0\fi
   \sectionnumber=0 \makechapterlabel}
\def\makechapterlabel{\let\sectionlabel=\relax
   \xdef\chapterlabel{\the\chapterstyle{\the\chapternumber}.}}
\def\alphabetic#1{\count255='140 \advance\count255 by #1\char\count255}
\def\Alphabetic#1{\count255='100 \advance\count255 by #1\char\count255}
\def\Roman#1{\uppercase\expandafter{\romannumeral #1}}
\def\roman#1{\romannumeral #1}
\def\Number#1{\number #1}
\def\BLANC#1{}
\def\titlestyle#1{\par\begingroup \interlinepenalty=9999
     \leftskip=0.02\hsize plus 0.23\hsize minus 0.02\hsize
     \rightskip=\leftskip \parfillskip=0pt
     \hyphenpenalty=9000 \exhyphenpenalty=9000
     \tolerance=9999 \pretolerance=9000
     \spaceskip=0.333em \xspaceskip=0.5em
     \iftwelv@\fourteenpoint\else\twelvepoint\fi
   \noindent #1\par\endgroup }
\def\spacecheck#1{\dimen@=\pagegoal\advance\dimen@ by -\pagetotal
   \ifdim\dimen@<#1 \ifdim\dimen@>0pt \vfil\break \fi\fi}
\def\TableOfContentEntry#1#2#3{\relax}
\def\chapter#1{\par \penalty-300 \vskip\chapterskip
   \spacecheck\chapterminspace
   \chapterreset \titlestyle{\chapterlabel\ #1}
   \TableOfContentEntry c\chapterlabel{#1}
   \nobreak\vskip\headskip \penalty 30000
   \wlog{\string\chapter\space \chapterlabel} }

\def\section#1{\par \ifnum\the\lastpenalty=30000\else
   \penalty-200\vskip\sectionskip \spacecheck\sectionminspace\fi
   \global\advance\sectionnumber by 1
   \xdef\sectionlabel{\the\sectionstyle\the\sectionnumber}
   \wlog{\string\section\space \sectionlabel}
   \TableOfContentEntry s\sectionlabel{#1}
   \noindent {\caps\enspace\sectionlabel\quad #1}\par
   \nobreak\vskip\headskip \penalty 30000 }
\def\subsection#1{\par
   \ifnum\the\lastpenalty=30000\else \penalty-100\smallskip \fi
   \noindent\undertext{#1}\enspace \vadjust{\penalty5000}}

\def\undertext#1{\vtop{\hbox{#1}\kern 1pt \hrule}}
\def\APPENDIX#1#2{\par\penalty-300\vskip\chapterskip
   \spacecheck\chapterminspace \chapterreset \xdef\chapterlabel{#1}
   \titlestyle{APPENDIX #2} \nobreak\vskip\headskip \penalty 30000
   \TableOfContentEntry a{#1}{#2}
   \wlog{\string\Appendix\ \chapterlabel} }
\def\Appendix#1{\APPENDIX{#1}{#1}}
\def\appendix{\APPENDIX{A}{}}
\def\unnumberedchapters{\let\makechapterlabel=\relax \let\chapterlabel=\relax
   \sectionstyle={\BLANC}\let\sectionlabel=\relax \sequentialequations }
%
%
%
\def\eqname#1{\relax \ifnum\equanumber<0
     \xdef#1{{\noexpand\rm(\number-\equanumber)}}%
       \global\advance\equanumber by -1
    \else \global\advance\equanumber by 1
      \xdef#1{{\noexpand\rm(\chapterlabel\number\equanumber)}} \fi #1}
\def\eqinsert#1{\noalign{\dimen@=\prevdepth \nointerlineskip
   \setbox0=\hbox to\displaywidth{\hfil #1}
   \vbox to 0pt{\kern 0.5\baselineskip\hbox{$\!\box0\!$}\vss}
   \prevdepth=\dimen@}}
%

%
%
\def\GENITEM#1;#2{\par \hangafter=0 \hangindent=#1
    \Textindent{$ #2 $}\ignorespaces}
\outer\def\newitem#1=#2;{\gdef#1{\GENITEM #2;}}
\newdimen\itemsize                \itemsize=30pt
\newitem\item=1\itemsize;
\newitem\sitem=1.75\itemsize;     
\newitem\ssitem=2.5\itemsize;     
\outer\def\newlist#1=#2&#3&#4;{\toks0={#2}\toks1={#3}%
   \count255=\escapechar \escapechar=-1
   \alloc@0\list\countdef\insc@unt\listcount     \listcount=0
   \edef#1{\par
      \countdef\listcount=\the\allocationnumber
      \advance\listcount by 1
      \hangafter=0 \hangindent=#4
      \Textindent{\the\toks0{\listcount}\the\toks1}}
   \expandafter\expandafter\expandafter
    \edef\c@t#1{begin}{\par
      \countdef\listcount=\the\allocationnumber \listcount=1
      \hangafter=0 \hangindent=#4
      \Textindent{\the\toks0{\listcount}\the\toks1}}
   \expandafter\expandafter\expandafter
    \edef\c@t#1{con}{\par \hangafter=0 \hangindent=#4 \noindent}
   \escapechar=\count255}
\def\c@t#1#2{\csname\string#1#2\endcsname}
\newlist\point=\Number&.&1.0\itemsize;
\newlist\subpoint=(\alphabetic&)&1.75\itemsize;
\newlist\subsubpoint=(\roman&)&2.5\itemsize;
%

%
%
%
%
\newcount\referencecount     \referencecount=0
\newcount\lastrefsbegincount \lastrefsbegincount=0
\newif\ifreferenceopen       \newwrite\referencewrite
\newif\ifrw@trailer
\newdimen\refindent     \refindent=30pt
\def\NPrefmark#1{\attach{\scriptscriptstyle [ #1 ] }}
\let\PRrefmark=\attach
\def\refmark#1{\relax\ifPhysRev\PRrefmark{#1}\else\NPrefmark{#1}\fi}
\def\refend@{\refmark{\number\referencecount}}
\def\refend{\refend@{}\space }
\def\refsend{\refmark{\count255=\referencecount
   \advance\count255 by-\lastrefsbegincount
   \ifcase\count255 \number\referencecount
   \or \number\lastrefsbegincount,\number\referencecount
   \else \number\lastrefsbegincount-\number\referencecount \fi}\space }
\def\refitem#1{\par \hangafter=0 \hangindent=\refindent \Textindent{#1}}
\def\Ref{\rw@trailertrue\REF}
\def\ref{\Ref\?}

\def\REF#1{\r@fstart{#1}%
   \rw@begin{\the\referencecount.}\rw@end}
\def\REFS#1{\r@fstart{#1}%
   \lastrefsbegincount=\referencecount
   \rw@begin{\the\referencecount.}\rw@end}
\def\r@fstart#1{\chardef\rw@write=\referencewrite \let\rw@ending=\refend@
   \ifreferenceopen \else \global\referenceopentrue
   \immediate\openout\referencewrite=referenc.txa
   \toks0={\catcode`\^^M=10}\immediate\write\rw@write{\the\toks0} \fi
   \global\advance\referencecount by 1 \xdef#1{\the\referencecount}}
{\catcode`\^^M=\active %
 \gdef\rw@begin#1{\immediate\write\rw@write{\noexpand\refitem{#1}}%
   \begingroup \catcode`\^^M=\active \let^^M=\relax}%
 \gdef\rw@end#1{\rw@@end #1^^M\rw@terminate \endgroup%
   \ifrw@trailer\rw@ending\global\rw@trailerfalse\fi }%
 \gdef\rw@@end#1^^M{\toks0={#1}\immediate\write\rw@write{\the\toks0}%
   \futurelet\n@xt\rw@test}%
 \gdef\rw@test{\ifx\n@xt\rw@terminate \let\n@xt=\relax%
       \else \let\n@xt=\rw@@end \fi \n@xt}%
}
\let\rw@ending=\relax
\let\rw@terminate=\relax
\let\splitout=\relax
\def\Closeout#1{\toks0={\catcode`\^^M=5}\immediate\write#1{\the\toks0}%
   \immediate\closeout#1}
%
%
\newcount\figurecount     \figurecount=0
\newcount\tablecount      \tablecount=0
\newif\iffigureopen       \newwrite\figurewrite
\newif\iftableopen        \newwrite\tablewrite
\def\FIG#1{\f@gstart{#1}%
   \rw@begin{\the\figurecount)}\rw@end}

\def\Fig{\rw@trailertrue\def\rw@ending{Fig.~\?}\FIG\?}
\def\fig{\rw@trailertrue\def\rw@ending{fig.~\?}\FIG\?}
\def\TABLE#1{\T@Bstart{#1}%
   \rw@begin{\the\tableecount:}\rw@end}
\def\Table{\rw@trailertrue\def\rw@ending{Table~\?}\TABLE\?}
\def\f@gstart#1{\chardef\rw@write=\figurewrite
   \iffigureopen \else \global\figureopentrue
   \immediate\openout\figurewrite=figures.txa
   \toks0={\catcode`\^^M=10}\immediate\write\rw@write{\the\toks0} \fi
   \global\advance\figurecount by 1 \xdef#1{\the\figurecount}}
\def\T@Bstart#1{\chardef\rw@write=\tablewrite
   \iftableopen \else \global\tableopentrue
   \immediate\openout\tablewrite=tables.txa
   \toks0={\catcode`\^^M=10}\immediate\write\rw@write{\the\toks0} \fi
   \global\advance\tablecount by 1 \xdef#1{\the\tablecount}}
%

%
%
%
\def\getfigure#1{\global\advance\figurecount by 1
   \xdef#1{\the\figurecount}\count255=\escapechar \escapechar=-1
   \edef\n@xt{\noexpand\g@tfigure\csname\string#1Body\endcsname}%
   \escapechar=\count255 \n@xt }
\def\g@tfigure#1#2 {\errhelp=\disabledfigures \let#1=\relax
   \errmessage{\string\getfigure\space disabled}}
\newhelp\disabledfigures{ Empty figure of zero size assumed.}
\def\figinsert#1{\midinsert\Tenpoint\medskip
   \count255=\escapechar \escapechar=-1
   \edef\n@xt{\csname\string#1Body\endcsname}
   \escapechar=\count255 \centerline{\n@xt}
   \bigskip\narrower\narrower
   \noindent{\it Figure}~#1.\quad }
%
%
%
\def\masterreset{\global\pagenumber=1 \global\chapternumber=0
   \global\equanumber=0 \global\sectionnumber=0
   \global\referencecount=0 \global\figurecount=0 \global\tablecount=0 }
\def\FRONTPAGE{\ifvoid255\else\vfill\penalty-20000\fi
      \masterreset\global\frontpagetrue
      \global\lastf@@t=0 \global\footsymbolcount=0}

\def\paperstyle{\letterstylefalse\normalspace\papersize}
\def\letterstyle{\letterstyletrue\singlespace\lettersize}
\def\papersize{\hsize=35 truepc\vsize=50 truepc\hoffset=2 truepc
               \skip\footins=\bigskipamount}
\def\lettersize{\hsize=5.5 truein\vsize=8.25 truein\hoffset=1.3 truein
	\voffset=1.25 truein
   \skip\footins=\smallskipamount \multiply\skip\footins by 3 }
\paperstyle   
%
%
\def\MEMO{\letterstyle \letterinfo={\hfil } \let\rule=\memorule
	\FRONTPAGE \memohead }
\let\memohead=\relax

\def\memit@m#1{\smallskip \hangafter=0 \hangindent=.1in
      \Textindent{\caps #1}}
\def\subject{\memit@m{Subject:}}
\def\topic{\memit@m{Topic:}}
\def\from{\memit@m{From:}}
\def\to{\relax \ifmmode \rightarrow \else \memit@m{To:}\fi }
\def\memorule{\medskip\hrule height 1pt\bigskip}
\newwrite\labelswrite
\newtoks\rw@toks

\def\addressee#1{\null\vskip .5truein\line{
\hskip 0.5\hsize minus 0.5\hsize\the\date\hfil}\bigskip
   \ialign to\hsize{\strut ##\hfil\tabskip 0pt plus \hsize \cr #1\crcr}
   \writelabel{#1}\medskip\par\noindent}
\def\rwl@begin#1\cr{\rw@toks={#1\crcr}\relax
   \immediate\write\labelswrite{\the\rw@toks}\futurelet\n@xt\rwl@next}
\def\rwl@next{\ifx\n@xt\rwl@end \let\n@xt=\relax
      \else \let\n@xt=\rwl@begin \fi \n@xt}
\let\rwl@end=\relax
\def\writelabel#1{\immediate\write\labelswrite{\noexpand\labelbegin}
     \rwl@begin #1\cr\rwl@end
     \immediate\write\labelswrite{\noexpand\labelend}}
\newbox\FromLabelBox

\let\labelend=\relax
\def\labelbegin#1\labelend{\setbox0=\vbox{\ialign{##\hfil\cr #1\crcr}}
     \MakeALabel }
\newtoks\FromAddress
\FromAddress={}
\def\MakeFromBox#1{\global\setbox\FromLabelBox=\vbox{\Tenpoint
     \ialign{##\hfil\cr #1\the\FromAddress\crcr}}}
\newdimen\labelwidth		\labelwidth=6in
\def\MakeALabel{\vskip 1pt \hbox{\vrule \vbox{
	\hsize=\labelwidth \hrule\bigskip
	\leftline{\hskip 1\parindent \copy\FromLabelBox}\bigskip
	\centerline{\hfil \box0 } \bigskip \hrule
	}\vrule } \vskip 1pt plus 1fil }
\newskip\signatureskip       \signatureskip=30pt
\def\signed#1{\par \penalty 9000 \medskip \dt@pfalse
  \everycr={\noalign{\ifdt@p\vskip\signatureskip\global\dt@pfalse\fi}}
  \setbox0=\vbox{\singlespace \ialign{\strut ##\hfil\crcr
   \noalign{\global\dt@ptrue}#1\crcr}}
  \line{\hskip 0.5\hsize minus 0.5\hsize \box0\hfil} \medskip }
\newbox\letterb@x
\def\lettertext{\par\unvcopy\letterb@x\par}
\def\multiletter{\setbox\letterb@x=\vbox\bgroup
      \everypar{\vrule height 1\baselineskip depth 0pt width 0pt }
      \singlespace \topskip=\baselineskip }
\def\letterend{\par\egroup}
%
%
%
\newskip\frontpageskip
\newtoks\Pubnum
\newtoks\pubtype
\newif\ifp@bblock  \p@bblocktrue
\def\PH@SR@V{\doubl@true \baselineskip=24.1pt plus 0.2pt minus 0.1pt
             \parskip= 3pt plus 2pt minus 1pt }
\def\PHYSREV{\paperstyle\PhysRevtrue\PH@SR@V}
\def\titlepage{\FRONTPAGE\paperstyle\ifPhysRev\PH@SR@V\fi
   \ifp@bblock\p@bblock \else\hrule height\z@ \relax \fi }
\def\nopubblock{\p@bblockfalse}

\frontpageskip=12pt plus .5fil minus 2pt
\pubtype={\tensl Preliminary Version}
\Pubnum={}
\def\p@bblock{\begingroup \tabskip=\hsize minus \hsize
   \baselineskip=1.5\ht\strutbox \topspace-2\baselineskip
   \halign to\hsize{\strut ##\hfil\tabskip=0pt\crcr
       \the\Pubnum\crcr\the\date\crcr\the\pubtype\crcr}\endgroup}
\def\title#1{\vskip\frontpageskip \titlestyle{#1} \vskip\headskip }
\def\author#1{\vskip\frontpageskip\titlestyle{\twelvecp #1}\nobreak}

\def\address#1{\par\kern 5pt\titlestyle{\twelvepoint\it #1}}
\def\andaddress{\par\kern 5pt \centerline{\sl and} \address}

\def\abstract{\par\dimen@=\prevdepth \hrule height\z@ \prevdepth=\dimen@
   \vskip\frontpageskip\centerline{\fourteenrm ABSTRACT}\vskip\headskip }

%
%
%

\def\\{\relax \ifmmode \backslash \else {\tt\char`\\}\fi }
\def\sequentialequations{\relax\if\equanumber<0\else\global\equanumber=-1\fi}

\def\journal#1&#2(#3){\unskip, \sl #1\unskip~\bf\ignorespaces #2\rm (19#3),}

\def\topspace{\hrule height 0pt depth 0pt \vskip}

\def\Buildrel#1\under#2{\mathrel{\mathop{#2}\limits_{#1}}}
\def\becomes#1{\mathchoice{\becomes@\scriptstyle{#1}}{\becomes@\scriptstyle
   {#1}}{\becomes@\scriptscriptstyle{#1}}{\becomes@\scriptscriptstyle{#1}}}
\def\becomes@#1#2{\mathrel{\setbox0=\hbox{$\m@th #1{\,#2\,}$}%
	\mathop{\hbox to \wd0 {\rightarrowfill}}\limits_{#2}}}

\let\int=\intop         
\def\lsim{\mathrel{\mathpalette\@versim<}}
\def\gsim{\mathrel{\mathpalette\@versim>}}
\def\@versim#1#2{\vcenter{\offinterlineskip
	\ialign{$\m@th#1\hfil##\hfil$\crcr#2\crcr\sim\crcr } }}
\def\big#1{{\hbox{$\left#1\vbox to 0.85\b@gheight{}\right.\n@space$}}}
\def\Big#1{{\hbox{$\left#1\vbox to 1.15\b@gheight{}\right.\n@space$}}}
\def\bigg#1{{\hbox{$\left#1\vbox to 1.45\b@gheight{}\right.\n@space$}}}
\def\Bigg#1{{\hbox{$\left#1\vbox to 1.75\b@gheight{}\right.\n@space$}}}
%
%
%
\let\sec@nt=\sec
\def\sec{\relax\ifmmode\let\n@xt=\sec@nt\else\let\n@xt\section\fi\n@xt}
\def\obsolete#1{\message{Macro \string #1 is obsolete.}}
\def\firstsec#1{\obsolete\firstsec \section{#1}}
\def\firstsubsec#1{\obsolete\firstsubsec \subsection{#1}}
\def\thispage#1{\obsolete\thispage \global\pagenumber=#1\frontpagefalse}
\def\thischapter#1{\obsolete\thischapter \global\chapternumber=#1}
\def\REFSCON{\obsolete\REFSCON\REF}
\def\splitout{\obsolete\splitout\relax}
\def\prop{\obsolete\prop \propto }
\def\nextequation#1{\obsolete\nextequation \global\equanumber=#1
   \ifnum\the\equanumber>0 \global\advance\equanumber by 1 \fi}
\def\BOXITEM{\afterassigment\B@XITEM\setbox0=}
\def\B@XITEM{\par\hangindent\wd0 \noindent\box0 }
\def\phyzzx{PHY\setbox0=\hbox{Z}\copy0 \kern-0.5\wd0 \box0 X}
%
%
\everyjob{\xdef\today{\monthname\ \number\day, \number\year}}
        
%

%

\hsize=6truein
\def\TITLEPAGE{\frontpagetrue}
\def\CALT#1{\hbox to\hsize{\tenpoint \baselineskip=12pt
	\hfil\vtop{\hbox{\strut CALT-68-#1}
	\hbox{\strut DOE RESEARCH AND}
	\hbox{\strut DEVELOPMENT REPORT}}}}

\def\CALTECH{\smallskip
	\address{California Institute of Technology, Pasadena, CA 91125}}

\def\AUTHOR#1{\vskip .5in \centerline{#1}}

\def\ABSTRACT#1{\vskip .5in \vfil \centerline{\twelvepoint \bf Abstract}
	#1 \vfil}
\def\ENDTITLEPAGE{\vfil\eject\pageno=1}

\def\sqr#1#2{{\vcenter{\hrule height.#2pt
      \hbox{\vrule width.#2pt height#1pt \kern#1pt
        \vrule width.#2pt}
      \hrule height.#2pt}}}

\def\section#1#2{
\noindent\hbox{\hbox{\bf #1}\hskip 10pt\vtop{\hsize=5in
\baselineskip=12pt \noindent \bf #2 \hfil}\hfil}
\medskip}

\def\underwig#1{	
	\setbox0=\hbox{\rm \strut}
	\hbox to 0pt{$#1$\hss} \lower \ht0 \hbox{\rm \char'176}}

\def\bunderwig#1{	
	\setbox0=\hbox{\rm \strut}
	\hbox to 1.5pt{$#1$\hss} \lower 12.8pt
	 \hbox{\seventeenrm \char'176}\hbox to 2pt{\hfil}}

\def\MEMO#1#2#3#4#5{
\frontpagetrue
\centerline{\tencp INTEROFFICE MEMORANDUM}
\smallskip
\centerline{\bf CALIFORNIA INSTITUTE OF TECHNOLOGY}
\bigskip
\vtop{\tenpoint
\hbox to\hsize{\strut \hbox to .75in{\caps to:\hfil}\hbox to 3.8in{#1\hfil}
\quad\the\date\hfil}
\hbox to\hsize{\strut \hbox to.75in{\caps from:\hfil}\hbox to 3.5in{#2\hfil}
\hbox{{\caps ext-}#3\qquad{\caps m.c.\quad}#4}\hfil}
\hbox{\hbox to.75in{\caps subject:\hfil}\vtop{\parindent=0pt
\hsize=3.5in #5\hfil}}
\hbox{\strut\hfil}}}

\tolerance=10000
\hfuzz=5pt

\def\vslash{\rlap{/}v}
\TITLEPAGE
\CALT{1844}
\setbox0=\hbox{\strut DOE RESEARCH AND}
\hbox to \hsize{\hfil\vtop{\hbox{MIT--CTP\#2177}
\hbox to \wd0{\hfil}}}
\bigskip
\titlestyle {Chiral Perturbation Theory for $B \rightarrow D^*$ and $B
\rightarrow D$  Semileptonic Transition Matrix Elements at Zero Recoil
\foot{Work supported in part by the U.S. Dept. of Energy
under Contract numbers DEAC-03-81ER40050 and DE-AC02-76ER03069 and by the
Texas National Research Laboratory Commission under Grant no. RGFY26C6.}}
\AUTHOR{Lisa Randall\foot{National Science Foundation Young Investigator
Award, Alfred P. Sloan Foundation Research Fellowship, Department of
Energy Outstanding Junior Investigator Award.}}
\bigskip
\centerline{{\it Massachusetts Institute of Technology, Cambridge, MA
02139}}
\AUTHOR{Mark B. Wise}
\CALTECH
\ABSTRACT{Heavy quark symmetry predicts the value of $B \rightarrow D$
and $B \rightarrow D^*$ transition matrix elements of the current $\bar
c \gamma_\mu (1 - \gamma_5)b$, at zero recoil (where in the rest frame of
the $B$ the $D$ or $D^*$ is also at rest).  We use chiral perturbation
theory to compute the leading
 corrections to these predictions which are generated
at low momentum, below the chiral symmetry breaking scale.}
\ENDTITLEPAGE

\eject

The interactions of a heavy quark $Q$ (i.e., $m_Q \gg \Lambda_{QCD}$)
are simplified by going over to an effective theory where the heavy
quark mass goes to infinity with its four-velocity fixed.  The effective
theory reveals a heavy quark spin-flavor symmetry$^{[1,2]}$ that is not
manifest in the full
theory of QCD.  Heavy quark symmetry has been used to predict many
properties of hadrons containing a single heavy quark.  For example, it
implies that all the form factors for $B \rightarrow De\bar \nu_e$ and
$B \rightarrow D^* e\bar\nu_e$ can  be expressed in terms of a single
universal function$^{[1]}$ and that the value of this function at zero recoil
(where in the rest frame of the $B$ the $D$ or $D^*$ is also at rest) is
known.$^{[1,3,4]}$  One of the most significant applications of
these results will be
determining the value of the Cabibbo--Kobayashi--Maskawa matrix element
$V_{cb}$.

It has been shown$^{[5]}$ that at zero recoil the matrix elements
which are necessary to determine $V_{cb}$ are known
up to corrections at order $1/m_Q^2$. These corrections have
been estimated in various models; dimensional analysis implies
the corrections will be of order $(\Lambda/m_Q)^2$, where $\Lambda$
is determined by the chiral symmetry breaking scale$^{[6]}$.
In addition to these ``matching" corrections to the current, there
are also corrections which are generated at low momentum that have a
nonanalytic dependence on the pion mass and on the
heavy quark suppressed mass difference between heavy meson vector
and pseudoscalar states.  It is these we compute in this letter.
We first  review the heavy meson chiral lagrangian and the
standard analysis of the heavy quark currents.  We use  chiral
perturbation theory to compute the dominant corrections at zero recoil.
Because the inclusion of kaons in the heavy meson chiral lagrangian is
suspect$^{[7]}$, we focus on  pion loop corrections.

The ground state heavy mesons with $Q\bar q_a$ flavor quantum numbers
(Here $a = 1,2$ and $q_1 = u, q_2 = d$) have $s_\ell^{\pi_{\ell}} =
{1\over 2}^-$, for the spin parity of the light  degrees of freedom.
Combining the spin of the light degrees of freedom with the spin of the
heavy quark gives (in the $m_Q \rightarrow \infty$ limit) two degenerate
doublets consisting of spin zero and spin-one mesons  that are denoted by
$P_a$ and  $P_a^*$ respectively.  In the case $Q = c, P_a = (D^0,
D^+)$ and $P_a^* = (D^{*0}, D^{*+})$ while for $Q = b, P_a = (B^-, \bar
B^0$) and $P_a^* = (B^{*-}, \bar B^{*0})$.  It is convenient to combine
the fields $P_a$ and $P_{a\mu}^*$ that destroy these mesons $(v^\mu
P_{a\mu}^* = 0)$ into a $4\times 4$ matrix $H_a$ given by$^{[8]}$
$$	H_a = \left({1 + \vslash\over 2}\right) (P_{a\mu}^* \gamma^\mu -
P_a \gamma_5) \,\, . \eqno (1)$$
(This is a compressed notation.  In situations where the type of
heavy quark $Q$ and its four-velocity $v$ are important  the $4\times 4$
matrix is denoted by $H_a^{(Q)} (v)$).  It transforms under the heavy
quark spin symmetry group $SU(2)_v$ as
$$	H_a \rightarrow S H_a \,\, , \eqno (2)$$
where $S\epsilon SU(2)_v$ and under Lorentz transformations as
$$ 	H_a \rightarrow D(\Lambda) H_a D(\Lambda)^{-1} \,\, , \eqno (3)$$
where $D(\Lambda)$ is an element of the $4\times 4$ matrix
representation of the Lorentz group.  It is also useful to introduce
$$	\eqalign{\bar H_a &= \gamma^0 H_a^{\dag} \gamma^0\cr
&= (P_{a\mu}^{*\dag} \gamma^\mu + P_a^{\dag} \gamma_5) {(1 + \vslash)\over 2}
\,\, .\cr} \eqno (4)$$
For $\bar H_a$ the transformation laws corresponding to these in eqs.
(2) and (3) become $\bar H_a \rightarrow \bar H_a S^{-1}$ and $\bar H_a
\rightarrow D(\Lambda) \bar H_a D(\Lambda)^{-1}$.

The strong interactions also have an approximate $SU(2)_L \times
SU(2)_R$ chiral symmetry that is spontaneously broken to the vector
$SU(2)_V$ isospin subgroup.   This
symmetry arises because the light up and down quarks have masses that
are small compared with the typical scale of the strong interactions.  (If
the strange quark is also treated as
light the chiral symmetry group becomes $SU(3)_L \times SU(3)_R$).
Associated with the spontaneous breaking of $SU(2)_L \times SU(2)_R$
chiral symmetry are the pions.  The low momentum strong interactions of
these pseudo Goldstone bosons are described by a chiral Lagrangian that
contains the most general interactions consistent with chiral symmetry.
The effects of the up and down quark masses are included by adding terms
that transform in the same way under chiral symmetry as the quark mass
terms in the QCD Lagrangian.

The pions are incorporated in a $2\times 2$ unitary matrix
$$	 \Sigma = \exp \left({2iM\over f}\right) \eqno (5)$$
where
$$	M = \left[\matrix{\pi^0/\sqrt{2} & \pi^+\cr
\pi^- & -\pi^0/\sqrt{2}\cr}\right] \,\, , \eqno (6)$$
and $f \simeq 132$ MeV is the pion decay constant.  Under a chiral
$SU(2)_L \times SU(2)_R$ transformation
$$	\Sigma \rightarrow L\Sigma R^{\dag}\,\, , \eqno (7)$$
where $L\epsilon SU(2)_L$ and $R\epsilon SU(2)_R$.  It is convenient
when discussing the interactions of the $\pi$ mesons with the $P_a$ and
$P_a^*$ mesons to introduce
$$	\xi = \exp \left({iM\over f}\right) \,\, . \eqno (8)$$
Under a chiral $SU(2)_L \times SU(2)_R$ transformation
$$	\xi \rightarrow L\xi U^{\dag} = U\xi R^{\dag}\,\, , \eqno (9)$$
where typically the special unitary matrix $U$ is a complicated
nonlinear function of $L, R$ and the pion fields.  However, for
transformations $V = L = R$ in the unbroken subgroup $U = V$.  We assign
the heavy meson fields the transformation law
$$	H_a \rightarrow H_b U_{ba}^{\dag} \,\, , \eqno (10)$$
under chiral $SU(2)_L \times SU(2)_R$ (In eq. (10) and for the remainder
of this paper repeated subscripts $a$ and $b$ are summed over $1,2$).

The low momentum strong interactions of pions with heavy $P_a$ and
$P_a^*$ mesons are described by the effective Lagrange
density$^{[9,10,11]}$
$$	{\cal L} = - i Tr \bar H_a v_\mu \partial^\mu H_a +  {1\over 2}
iTr \bar H_a H_b v^\mu (\xi^{\dag} \partial_\mu \xi + \xi \partial_\mu
\xi^{\dag})_{ba}$$
$$	+ {1\over 2} ig Tr \bar H_a \gamma_\nu \gamma_5 H_b(\xi^{\dag}
\partial^{\nu} \xi - \xi \partial^\nu \xi^{\dag})_{ba} + ... \,\, , \eqno
(11)$$
where the ellipsis denote terms with more derivatives.  This Lagrange
density is the most general one invariant under $SU(2)_L \times
SU(2)_R$ chiral symmetry, heavy quark spin symmetry, parity and Lorentz
transformations.  Heavy quark flavor symmetry implies that $g$ is
independent of the heavy quark mass.  Note that in eq. (11) factors of
$\sqrt{m_P}$ and $\sqrt{m_{P^{*}}}$ have been absorbed into the $P_a$
and $P_a^{*\mu}$ fields so they have dimension ${3\over 2}$.

The coupling $g$ determines the $D^{*+} \rightarrow D^0 \pi^+$ decay
width
$$	\Gamma (D^{*+} \rightarrow D^0 \pi^+) = {1\over 6\pi} ~
{g^2\over f^2} |\vec p_\pi |^3\,\, . \eqno (12)$$
Using the measured branching ratio, for this decay$^{[12]}$ and the recent
limit on the $D^*$ width$^{[13]}$ gives the bound $g^2 \leq 0.5$.

It is possible to include the symmetry breaking effects of order $m_q$
and $1/m_Q$ into the effective Lagrangian for pion heavy meson strong
interactions. Explicit chiral symmetry breaking effects are suppressed relative
to the leading correction which we calculate. We include
terms of order $m_q$
only through the nonzero pion mass.

The $1/m_Q$ terms that break the spin-flavor heavy quark symmetry give
rise to the additional term
$$	\delta {\cal L}^{(2)} = {\lambda_2\over m_Q} Tr \bar H_a
\sigma^{\mu\nu} H_a \sigma_{\mu\nu} + {\lambda'_2\over m_Q} Tr \bar H_a
H_a + ... \eqno (15)$$
in the Lagrange density.  The second term in eq. (15)
violates the heavy quark flavor symmetry but not the spin symmetry and
the first term violates both the heavy quark spin and flavor
symmetries.  The ellipsis denote terms with derivatives.  Included in
these are, for example, the $1/m_Q$ correction to $g$.
The second term in eq. (15) can be removed by phase transformations on
the heavy meson fields.  Therefore,
at  the leading order in chiral perturbation
theory, it is only the first term in eq. (15) that produces violations of
heavy quark spin-flavor symmetry in the
low energy heavy meson lagrangian. The
effect of  $\lambda_2$ is to shift the mass
of the pseudoscalar  relative to the  vector meson by
$$	\Delta^{(Q)} = m_{P^{*(Q)}} - m_{P^{(Q)}} \,\,  \eqno (16)$$
which distinguishes the heavy meson propagators.
Explicitly, $\Delta^{(Q)} = - 2 \lambda_2/m_Q$,
which determines $\lambda_2 \approx 300 {\rm MeV}$.
 Heavy quark mass suppressed
operators in which the pion couples can be neglected at leading
order in chiral perturbation theory.

Semileptonic decays $B \rightarrow D e\bar\nu_e$ and $B \rightarrow D^*
e\bar\nu_e$ can be studied using chiral perturbation theory.  It is the
hadronic matrix elements of $\bar c \gamma_\mu (1 - \gamma_5)b$ that are
needed for these decays.  This operator is a singlet under $SU(2)_L
\times SU(2)_R$ and in chiral perturbation theory its $B(v) \rightarrow
D(v')$ and $B(v) \rightarrow D^*(v')$ hadronic matrix elements are
given by those of
$$	\bar c \gamma_\mu (1 - \gamma_5)b = - \beta (v \cdot v') Tr \bar
H_a^{(c)} (v') \gamma_\mu (1 - \gamma_5) H_a^{(b)} (v) + ... \,\, ,
\eqno (17)$$
where the ellipsis denote terms with derivatives, factors of $m_q$ and
factors of $1/m_Q$.  Heavy quark symmetry implies the normalization
$$	\beta (1) = \left[{\alpha_s (m_b)\over \alpha_s
(m_c)}\right]^{-6/25} \,\, , \eqno (18)$$
at zero recoil.  The deviation of $\beta(1)$ from unity arises from
calculable perturbative QCD corrections associated with momentum
scales between the bottom and charm quark masses.  In eq. (18) these
perturbative corrections have been summed using the leading logarithmic
approximation$^{[14,15]}$

The zero recoil matrix elements
$$	<D^* (v,\epsilon) |\bar c \gamma_\mu \gamma_5 b| B(v)> = 2
\epsilon_\mu^* \beta (1) \eqno (19a)$$
and
$$	<D(v) |\bar c \gamma_\mu b| B(v)> = 2 v_\mu \beta (1) \eqno (19b)$$
receive no corrections at order $1/m_Q$.$^{[5]}$  This result (which is often
referred to as Luke's theorem) is important for extracting a precise
value of $V_{cb}$ from semileptonic $B$-decays.

In the chiral perturbation theory, the corrections to eqs. (19) arise
from the tree level matrix elements of  $1/m_Q^2$ suppressed
operators. The coefficients of these operators can be estimated, through
models or dimensional analysis.  In addition there are one-loop Feynman
diagrams that give nonanalytic dependence on the pion mass.  These
one-loop diagrams depend on the subtraction point $\mu$ and this
dependence is canceled by the coefficient of an operator whose
symmetry structure is that of the  spin symmetry breaking operator
inserted twice
$${\cal O}_s= {1\over m_c^2} Tr \sigma^{\alpha \beta} \sigma^{\mu \nu} \bar
H_a^{(c)} (v)\sigma^{\mu \nu}{(1 + \vslash)\over 2} \sigma^{\alpha \beta}
{(1 + \vslash)\over 2} \gamma_\mu (1 - \gamma_5) H_a^{(b)} (v)\,\, . \eqno
(20)$$
For simplicity in this letter we neglect effects suppressed by powers of the
bottom
quark mass.

Including the effects of the Feynman diagram in Fig. 1, one loop wave
function renormalization and a tree level ``counter term'' eq. (19b) becomes
$$	<D(v) |\bar c \gamma_\mu b| B(v)> = 2 v_\mu \beta (1) \Bigg\{1 +
C(\mu)/m_c^2 - {3g^2\over 2} \left({\Delta^{(c)}\over 4\pi f}\right)^2$$
$$	\left[\ell n (\mu^2/m_\pi^2) + f (\Delta^{(c)}/m_\pi)\right]\Bigg\}
\,\, , \eqno (21)$$
where
$$	f(x) = 2 \int_0^\infty dq {q^4\over (q^2 + 1)^{3/2}}
\left\{{1\over [(q^2 + 1)^{1/2} + x]^2} - {1\over q^2 + 1}\right\} \,\, .
\eqno (22)$$
The function $f(x)$ is given by
$$
f(x)={2 \over 3}+\pi \left[-{2 i} x^{-3}(x^2-1)^{3/2}-2 x^{-3}+3i
x^{-1}(x^2-1)^{1/2}\right]
$$
$$+ 4x^{-2} +2x^{-3}(x^2-1)^{3/2} \log \left({(x^2-1)^{1/2}-x\over
(x^2-1)^{1/2}+x}\right)$$
$$-3 x^{-1}(x^2-1)^{1/2} \log \left({(x^2-1)^{1/2}-x
\over (x^2-1)^{1/2}+x}
\right)\,\, . \eqno (23)$$
The expression for $f(x)$ is valid for all $x$. However, for $x<1$,
it is more sensible to replace the logarithm by an arctangent
and to rewrite the square roots so they are real. In the limit
of small $x$, $f(x)$
can be expanded in $x$
beginning at order $x$.
Notice that the loop contribution is proportional to ${\Delta^{(c)}}^2$
because we are calculating a quantity which is protected at order $1/m_Q$.
Had we calculated wave function or decay constant renormalization for example,
there are in general  contributions of order $\Delta^{(c)}$.

In eq. (21), $C(\mu)$ is the contribution of a tree level ``counter
term'' of
order $1/m_c^2$.  Its coefficient has subtraction point dependence that
cancels that in the logarithm.  For $\mu$ of order the chiral symmetry
breaking scale $C(\mu)$ contains no large logarithms and is, at least
formally, smaller than the term which is enhanced by the logarithm of the
pion mass.  The
function $f$ takes into account the effects of corrections of order
$(1/m_c)^{2 + n}, n = 1, 2, ...$.  It is enhanced by powers of
$(1/m_\pi)$ over terms we have neglected and so should provide a
reliable estimate of the $(1/m_c)^{2+n}, n=1,2,...$ effects.
Experimentally,
$\Delta^{(c)} = m_\pi$ and the expression in eq. (23) gives
$f(1)=2({7\over 3}-\pi )$. Numerically, for $g^2=0.5$ and $\mu =1GeV$, the
correction from the  logarithmically enhanced term
is $-2.1\%$ and the correction
from $f$ is $0.9\%$. Although the amplitude appears to be complex,
it can be checked that the imaginary part cancels.

Including the effects of the Feynman diagram in Fig. 1, one loop wave
function renormalization  and a tree level ``counter term'' eq. (19a) becomes
$$	<D^* (v,\epsilon)|\bar c \gamma_\mu \gamma_5 b| B(v)> = 2
\epsilon_\mu^* \beta (1) \Bigg\{ 1 + {C'(\mu)}/m_c^2 - {g^2\over 2}
\left({\Delta^{(c)}\over 4\pi f}\right)^2$$
$$	\left[ \ell n (\mu^2/m_\pi^2) + f ( -
\Delta^{(c)}/m_\pi)\right]\Bigg\} \,\, . \eqno (24)$$
Here,  the scale dependence of $C'(\mu)$ is 1/3 that of $C(\mu)$
but because there are two independent counterterms, the ``matching''
contribution is different from that of $C(\mu)/3$.
For this current, the one--loop calculation gives an imaginary
part to the amplitude because the intermediate pion and $D$ meson
states can be  simultaneously on shell.
 However, because $\Delta^{(c)}$ is very close to $m_\pi$,
this imaginary part is negligible and it is a good approximation is to
use eq. (23) with $\Delta^{(c)}$ set equal to the pion mass in $f$.  The
expression in
eq. (23) gives $f(-1) =2({7\over 3}+\pi )$. Numerically for $g^2=0.5$
and $\mu =1GeV$ the correction form the ``large logarithm'' is about
$-0.7\%$ and the correction from $f$ is about $-2.0\%$.

In this paper we have used chiral $SU(2)_L \times SU(2)_R$.  The
leading order contribution
of kaon and eta loops is absorbed into  $C(\mu)$ and
$C'(\mu)$.  These constants are independent of $m_c$ and $m_\pi$ and are
coefficients of two independent $1/m_c^2$ operators,
 one of which can be taken to be that in  eq. (20).
If we had used chiral $SU(3)_L \times SU(3)_R$ symmetry, the kaon and
eta loops would have given  a smaller contribution than the pion loops.

In this paper we have shown that corrections to the heavy quark symmetry
relations in eqs. (19) are computable using chiral perturbation theory.
For corrections of order $(1/m_Q)^{2+n}, n=1,2, ...$ the terms we have kept
are enhanced over those that were neglected by a factor of $1/m_{\pi }$ and
we have confidence that our calculation accurately takes into account
physical effects of this order. Because the pion mass occurs in the
denominator the expansion in powers of $1/m_Q$ breaks down in the limit
where the pion mass goes to zero. Since $m_{\pi }$ is about equal to
$\Delta^{(c)} $ for $Q=c$ all terms of order $(1/m_c)^{2+n}, n=1,2,...$ are
of comparable importance.

The factors of the pion mass in the denominator arise from low momentum in
the one loop Feynman diagrams. If one matches the heavy quark
theory onto the heavy meson chiral Lagrangian then (even for massless pions)
the coefficients of operators in the chiral Lagrangian (evaluated at a
subtraction point $\mu \simeq 1GeV$) have a well defined expansion in powers
of $1/m_Q$.

It is encouraging from the standpoint of the validity of the
heavy quark effective theory that the contribution we calculate
is not very large.  This is due in part to the factor of $g^2$ and
in part due to the partial cancellation between diagrams.
 However, because the one--loop result is small,
the contributions of $C(\mu)$ and $C'(\mu)$, although not logarithmically
enhanced, might nevertheless be comparable in magnitude.  These other
effects need to be estimated in phenomenological
models like QCD sum rules and the nonrelativistic constituent quark
model.$^{[16]}$
The very low
momentum effects we have explicitly computed
are not represented in such models.  Therefore, our results should be
added to their  predictions.

\noindent{\bf References}

\item{1.}  N. Isgur and M.B. Wise, Phys. Lett., {\bf B232} (1989) 113;
Phys. Lett., {\bf B237} (1990) 527.

\item{2.}  H. Georgi, Phys. Lett., {\bf B240} (1990) 447;  E. Eichten
and B. Hill, Phys. Lett., {\bf B234} (1990) 511.

\item{3.}  S. Nussinov and W. Wetzel, Phys. Rev., {\bf D36} (1987) 130.

\item{4.}  M.B. Voloshin and M.A. Shifman, Sov. J. Nucl. Phys., {\bf
47} (1988) 199.

\item{5.}  M. Luke, Phys. Lett., {\bf B252} (1990) 447.

\item{6.}  L. Randall and E. Sather, MIT-CTP \#2167 (1992) unpublished.

\item{7.}  L. Randall and E. Sather, MIT-CTP \#2166 (1992) unpublished.

\item{8.}  See for example; H. Georgi, Heavy Quark Effective Field
Theory, in Proceedings of the Theoretical Advanced Study Institute 1991,
ed. R.K. Ellis, C.T. Hill and J.D. Lykken, World Scientific (1992).

\item{9.}  M.B. Wise, Phys. Rev., {\bf D45} (1992) 2188.

\item{10.}  G. Burdman and J. Donoghue, Phys. Lett., {\bf B280} (1992)
287.

\item{11.}  T.M. Yan, et al., Phys. Rev., {\bf D46} (1992) 1148.

\item{12.}  F. Butler, et al., (CLEO collaboration) CLNS-92-1143 (1992)
unpublished.

\item{13.}  S. Barlag, et al., (ACCMOR collaboration) Phys. Lett., {\bf
B278} (1992) 480.

\item{14.}  M.B. Voloshin and M.A. Shifman, Sov. J. Nucl. Phys., {\bf
45} (1987) 292.

\item{15.}  H.D. Politzer and M.B. Wise, Phys. Lett., {\bf B206} (1988)
681; {\bf B208} (1988) 504.

\item{16.}  A.F. Falk and M. Neubert, SLAC-PUB-5897 (1992) unpublished.

\noindent {\bf Figure Caption}
\item {\rm Fig. 1} One loop Feynman diagram contributing to $B \to D^*$
and $B \to D$ transition matrix elements at zero recoil.  The dashed
line is a pion propagator and the shaded square is an insertion of the
current in eq. (17).

\bye